\documentstyle[aps,prl,epsf,floats,twocolumn]{revtex}
\begin{document}
\draft
\twocolumn[\hsize\textwidth\columnwidth\hsize\csname @twocolumnfalse\endcsname
\title{Mesoscopic Behavior Near a Two-Dimensional Metal-Insulator Transition}
\author{Dragana Popovi\'{c}$^{1}$ and S. Washburn$^{2}$}
\address{$^{1}$Dept. of Physics, City College of the City University of New 
York, New York, NY 10031 and \\ National High Magnetic Field Laboratory, 
Florida State University, Tallahassee, FL 32306 \\ $^{2}$Dept. of Physics and 
Astronomy, The University of North Carolina at Chapel Hill, Chapel Hill, NC 
27599}
\date{\today}
\maketitle

\begin{abstract}

We study conductance fluctuations in a two-dimensional electron gas 
as a function of chemical potential (or gate voltage) 
from the strongly insulating to the metallic regime.  
Power spectra of the fluctuations decay with two distinct exponents ($1/v_{l}$
and $1/v_{h}$).  For conductivity
$\sigma\sim 0.1~e^{2}/h$, we find a third exponent
($1/v_{i}$) in the shortest samples, and non-monotonic dependence of $v_{i}$
and $v_{l}$ on $\sigma$.  We study the dependence of $v_{i}$, $v_{l}$,
$v_{h}$, and the variances of corresponding
fluctuations on $\sigma$, sample size, and temperature.
The anomalies near $\sigma\simeq 0.1~e^{2}/h$ indicate that the dielectric
response and screening length are critically behaved, i.~e. that Coulomb
correlations dominate the physics.

\end{abstract}

%
%
%
%
%
]

The metal-insulator transition (MIT) is one of the fundamental
problems in condensed matter science.  Recent theoretical
work~\cite{Dobro} strongly suggests the crucial role played by 
electron-electron interactions in the transition regime.  However, since
the development of the scaling theory of localization~\cite{gang}, it has 
been asserted that all states are localized in 2D, in agreement
with early experiments~\cite{weakloc} on relatively low-mobility samples.
A recent experiment~\cite{Krav} on a two-dimensional electron gas (2DEG) in Si 
metal-oxide-semiconductor field-effect transistors (MOSFETs) provides evidence 
for the existence of a true MIT.  The samples used in that
experiment~\cite{Krav} had a much higher mobility and, as a result, 
the Coulomb interactions played a greater role relative to disorder.
Thus it has been speculated~\cite{Krav} that this MIT is driven by interaction
effects.  Using mesoscopic measurements, we provide direct evidence for the
crucial role of Coulomb interactions at the MIT in a 2DEG.

We investigate the statistics of conductance fluctuations in
a 2DEG as it undergoes a transition from strongly insulating to metallic
behavior.  In the insulating regime, electrons move in a strong, random
potential by tunneling through localized states.  In our relatively small 
samples and at low temperatures, the total number of states that 
contribute to conduction can be small, e.~g. of the order of 50--100.
By sweeping the gate voltage $V_{g}$, the chemical potential $\mu$ is shifted 
relative to the energy of localized states.  As a result, the number and
nature of localized states that dominate the transport change, and the 
conductance $G$ changes up to several orders of magnitude.  In addition, it is
well established that, deep in the insulating regime, the density of states 
$D(E)$ increases exponentially with increasing $V_{g}$~\cite{AFS}.

Our measurements were carried out on n-channel MOSFETs fabricated
on the (100) surface of Si doped at $\approx 3\times 
10^{14}$~acceptors/cm$^{3}$ with 500~\AA\,\, gate oxide thickness and 
(unintentional) oxide charge $< 10^{10}$~cm$^{-2}$.  The peak mobilities of 
our samples were of the order of 2~m$^{2}$/Vs -- comparable to those 
exhibiting a true MIT~\cite{Krav}.  In contrast to those samples, ours are 
much smaller: rectangular
with source-to-drain lengths $L=1-8~\mu$m, and widths $W=11.5-162~\mu$m.  
They were short enough to exhibit conductance fluctuations,
and wide enough to ensure good statistics.  Measurements were performed in a
dilution refrigerator with a lock-in at 
$\sim 10$~Hz and an excitation voltage of $0.2~\mu$V rms.
Conductance was measured as a function
of $V_{g}$ at temperatures $0.01< T < 0.8$~K.  

The behavior of the typical conductance, i.~e. of $\langle\ln G\rangle$ 
(averaging over intervals of $V_{g}$), in these samples
has been studied in detail~\cite{Pop}.  At very low $V_{g}$ and
for $T< 0.1$~K, the conduction is due to resonant tunneling through channels
containing a few localized states.  At higher $T$, the conduction proceeds via
variable-range hopping along isolated chains of several hops, consistent with
the model in Ref.~\cite{glazman}.  The conductance
of each channel depends not only on the energy and position of each
localized state in the channel but also on the Coulomb interactions between 
electrons.  At higher $V_{g}$, the current paths become more complicated
as more and more states contribute to conduction.

\begin{figure}[t]
\epsfxsize=3.6in \epsfbox{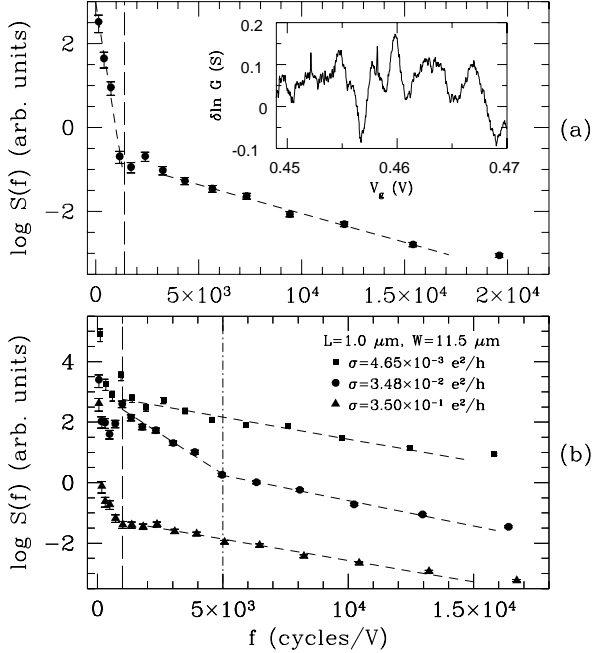}\vspace{5pt}
\caption{Power spectrum $S(f)$ of fluctuations $\delta\ln G$. (a) $L=2.0~\mu$m,
$W=162~\mu$m; $T=0.023$~K.  The vertical dashed line separates the 
low-frequency and high-frequency regions.  The fluctuations $\delta\ln G$ vs. 
$V_{g}$, corresponding to this $S(f)$, are shown in the inset.
$\sigma =1.37\times 10^{-2}e^{2}/h$.
(b) $T=0.045$~K.  The vertical dashed line 
separates the low-frequency region from the rest of the spectrum at
approximately the same frequency $f_{0}$ as in (a).  A new energy scale 
emerges at intermediate frequencies (located between the two vertical lines in
the plot) in the transition regime between strongly localized and metallic 
behavior.  The correlation voltage at the highest frequencies remains 
unchanged.  
\label{specs2}}
\end{figure}

\begin{figure}
\epsfxsize=3.2in \epsfbox{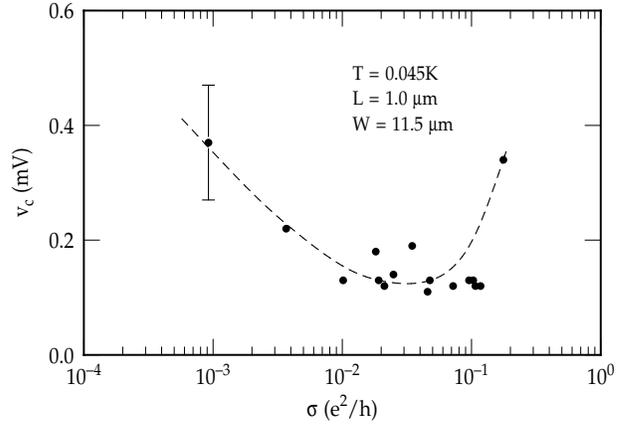}\vspace{6pt}
\caption{Characteristic voltage scale $v_{c}=1/f_{c}$, corresponding to the 
boundary $f_{c}$ between the intermediate-frequency and highest-frequency 
regions of the spectrum, as a function of $\sigma$.  
The dashed line is a guide to the eye.
\label{third}}
\end{figure}

The conductance fluctuates
on two, and sometimes three different scales of $V_{g}$~\cite{Pop}.
These scales are apparent in the raw data (see Fig.~1(a) inset), but a detailed
statistical analysis is necessary for inferring the underlying physics.
We have analyzed the
power spectra of fluctuations $\delta\ln G=\ln G(V_{g})-\langle\ln 
G(V_{g})\rangle$ for different $V_{g}$ ranges or conductivities 
$\sigma=(L/W)\exp \langle\ln G\rangle$,
sample sizes, and temperatures.  The power spectrum is a 
Fourier transform of the autocorrelation function 
$C(\Delta V_{g})=\langle\delta\ln G(V_{g})\delta\ln G(V_{g}+\Delta 
V_{g})\rangle$.  A
typical power spectrum $S(1/\Delta V_{g})$ is presented in 
Fig.~\ref{specs2}(a).  There
are two distinct regions of exponential 
decay, $S(f)=S(0)\exp (-2\pi vf)$, each corresponding to a Lorentzian
$C(\Delta V_{g})$ with characteristic widths $\sim v$.
The voltage correlation scales $v_{l}$ and $v_{h}$ 
represent the average peak spacings of the corresponding fluctuations.  
They are obtained by fitting both the low-frequency ($f<f_{0}$) and
high-frequency ($f_{0}<f$)
parts of $S(f)$ separately by the above exponential form.  
$f_{0}$, which separates the low-frequency part from the rest of the spectrum,
does not depend on any of the parameters within the scatter of our data;
$v_{0}=1/f_{0}=(0.45\pm 0.15)$~mV, averaged over all of our 
measurements.  Fig.~\ref{specs2}(b) illustrates the evolution of 
the spectrum as $\sigma$ increases.  While the correlation voltage
at the highest frequencies remains essentially unchanged 
($\sim 50~\mu$V) over a wide range of $\sigma$, we observe the 
emergence of a new distinct correlation voltage $v_{i}$ (i.~e. a new
Lorentzian in $C(\Delta V_{g})$) at intermediate 
frequencies.  From Fig.~\ref{third}, which plots the location of the 
boundary $f_{c}$ between the intermediate-frequency and the 
highest-frequency regions (shown by the dash-dotted vertical line in 
Fig.~\ref{specs2}(b)), we see that this new feature of
$S(f)$ is most pronounced for $10^{-2}<\sigma~(e^{2}/h)<10^{-1}$.
{\em It vanishes when the system becomes either more insulating or more 
metallic.} 

>From the work on quantum dots and wires~\cite{CB}, and tunneling through
localized states~\cite{Savchenko}, it is known that Coulomb interactions 
give rise to peaks in conductance such that the peak spacing is determined by
the strength of the interaction.  Therefore, at least some of the measured
correlation voltages should be proportional to the typical Coulomb energy in
the system.  As the MIT is approached from the insulating side, the Coulomb 
energy $U$ in our samples increases as a result of a decrease in the mean
separation of the electrons.  As the MIT is approached from the metallic side,
$U$ also increases because the screening becomes less 
efficient~\cite{AFS,Gleb,Lee}.
Therefore, we expect $U$ to have a {\em maximum} in the transition region.  

\begin{figure}[t]
\epsfxsize=3.4in \epsfbox{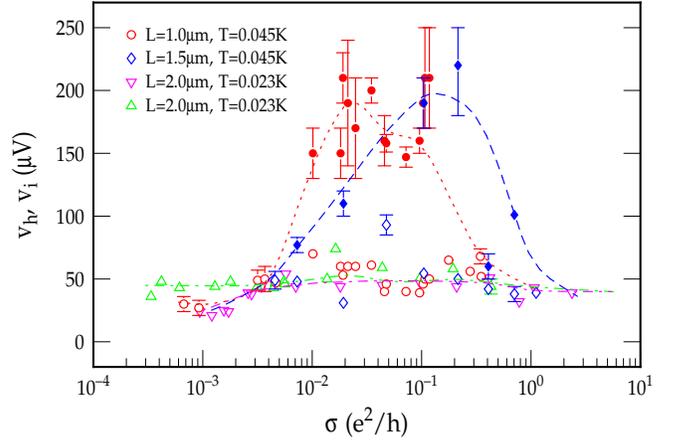}\vspace{6pt}
\caption{High-frequency ($f_{0}<f$) correlation voltage vs. $\sigma$ for 
samples of different size.  $v_{i}$ ($f_{0}<f<f_{c}$) is shown with solid
symbols, and $v_{h}$ ($f_{c}<f$) with open symbols.  One sample 
($\bigtriangleup$) had $W=162~\mu$m, and the rest had $W=11.5~\mu$m.
Dashed lines guide the eye.
\label{vh}}
\end{figure}

\begin{figure}
\epsfxsize=3.2in \epsfbox{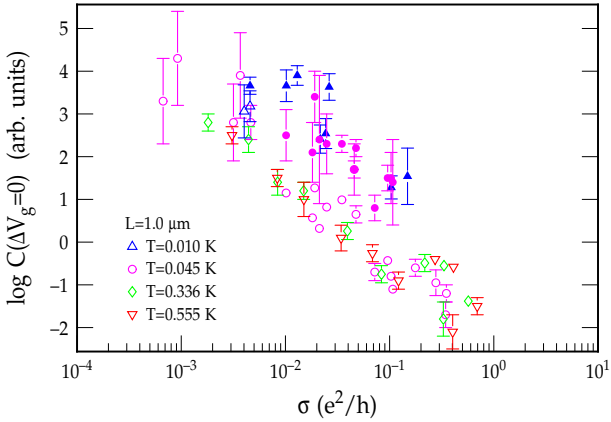}\vspace{6pt}
\caption{Variance $C(0)$ of high-frequency ($f_{0}<f$) fluctuations
vs. $\sigma$ at several temperatures.  Solid symbols are
from the $f_{0}<f<f_{c}$ part of the spectrum, and open symbols from the
$f_{c}<f$ region.  $W=11.5~\mu$m.
\label{var}}
\end{figure}

Obviously, it is important to analyze the evolution of the correlation voltages
with $\sigma$ in more detail.  We start with the high-frequency 
($f_{0}<f$) part.  Fig.~\ref{vh} presents $v_{h}$ and $v_{i}$ for
several samples of different size.  Fits to the decay rate are shown for
$f_{0}<f<f_{c}$ (solid symbols) and for $f_{c}<f$ (open symbols).  For
$f_{c}<f$, this correlation voltage $v_{h}$ remains of the order of 
$\sim 50~\mu$V, independent of sample size, $T$, and even $\sigma$.
For the shortest samples, however,  the third correlation scale $v_{i}$
(solid symbols in Figs.~\ref{vh} and \ref{var}) appears near $\sigma\simeq
0.1~e^{2}/h$. $v_{i}$ is observable only in the smallest samples, and it also
disappears at higher $T$, so it seems that this feature of $S(f)$ vanishes
as a result of averaging.  $v_{i}$ goes through a dramatic (increasing by
400\%) maximum near $\sigma =0.1~e^{2}/h$.

In order to characterize the fluctuations completely, we have also analyzed
their variances $C(\Delta V_{g}=0)$.  Fig.~\ref{var} shows how $C(0)$ of the 
fluctuations for $f_{0}<f$ depends on $\sigma$ in our smallest
sample.  For both types of fluctuations ($f_{0}<f<f_{c}$ and $f_{c}<f$) in
the high-frequency range, we observe a power-law decrease of $C(0)$ with an
increasing $\sigma$.  For the highest-frequency fluctuations, we find
no $T$-dependence of the variance in any of our samples but we have evidence 
that $C(0)$ is reduced by increasing the sample area.  Although the variance of
the intermediate-frequency fluctuations seems to decay with increasing $T$,
we do not have sufficient data to determine the functional form.  For this 
part of the spectrum, $C(0)$ seems to be suppressed rapidly by an increase of 
sample area.

\begin{figure}
\epsfxsize=3.2in \epsfbox{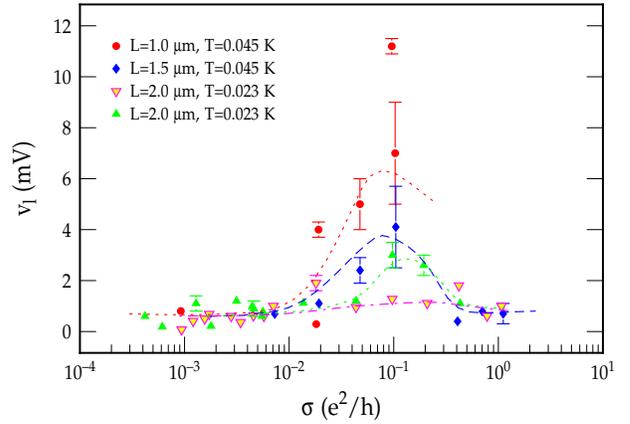}\vspace{6pt}
\caption{Correlation voltage $v_{l}$ of low-frequency ($f<f_{0}$) fluctuations
vs. $\sigma$ for samples of different size.  One sample ($\bigtriangleup$) had
$W=162~\mu$m, and the rest had $W=11.5~\mu$m.  The lines are guides to the eye.
\label{vl}}
\end{figure}

In the low-frequency range ($f<f_{0}$), the correlation
scale $v_{l}$ has a {\em maximum} as a function of $\sigma$ (Fig.~\ref{vl}),
in a qualitative agreement with the expected behavior of $U$.
The maximum occurs again for $10^{-2}<\sigma~(e^{2}/h)<10^{-1}$.
It is suppressed by increasing the sample size.  The results obtained on two
$L=2~\mu$m samples with different $W$ point to the absence of area averaging
in $v_{l}$.  A rise
in temperature shifts the peak to higher values of $\sigma$.  For
example, in our smallest sample ($L=1.0~\mu$m and $W=11.5~\mu$m) the peak moves
from $\sigma\approx 2.5\times 10^{-2}~(e^{2}/h)$ at $T=0.010$~K to
$\sigma\approx 2.0\times 10^{-1}~(e^{2}/h)$ at $T=0.555$~K.  It should
also be noted that $v_{l}$ is about 1--2 orders of magnitude larger than
$v_{h}$.  $C(0)$ for low-frequency fluctuations does not depend on any of the
parameters that we have varied except that there seems to be a small decrease
of $C(0)$ as $\sigma$ increases beyond $10^{-1}~(e^{2}/h)$.  

The rate
of the chemical potential shift with $V_{g}$ is~\cite{AFS} $d\mu/dV_{g}=
(d\mu/dn_{s})(dn_{s}/dV_{g})=(1/D(E))(C/e)$, where $n_{s}$ is the 2D electron
density, and $C$ is the gate capacitance of
our devices ($6.92\times 10^{-4}$~F/m$^{2}$).  Since it is not possible to
relate $D(E)$ to $\sigma$ without an adequate theoretical model of
transport, we are not able to obtain, for example, an accurate dependence of 
various correlation energies on $\sigma$.  It is possible, however, to
make some estimates, assuming that $D(E)$ varies from its constant 2D value
of $1.6\times 10^{14}$~cm$^{-2}$eV$^{-1}$ in the metallic regime down to about
10--20\% of that value in the insulating limit.  

With this in mind, we find that the voltage correlation scale of 
$\sim 50~\mu$V, corresponding to the highest-frequency regime, varies from 
about 1.35~$\mu$eV
at low $\sigma$, to $0.15~\mu$eV at the highest $\sigma$.  In spite
of a relatively weak dependence of correlation energy on $\sigma$, these
fluctuations disappear eventually at high $V_{g}$ because of a decrease in
their amplitude.  It is interesting that these fluctuations are well resolved 
all the way up to 0.8~K even though their correlation energy is much smaller 
than $k_{B}T$ which, while not intuitively plausible at first glance, is quite
reasonable upon detailed consideration of this issue for a related 
problem~\cite{Efetov}.
Given that this is the smallest energy scale, hardly dependent
on any of the parameters, and that these fluctuations are so ubiquitous, we are
led to speculate that their correlation energy is related to the energy level
spacing in our system.  If true, it would correspond to a sample area of about
$2\times 2~\mu$m.  This is somewhat smaller than the area of our devices but it
is consistent with the fact that only a fraction of the sample dominates 
conduction, as discussed in the introduction.  We recognize that there are
other energy scales of importance in this problem, but without a detailed model
for the transport, we must forgo further speculation.

In the same fashion, we find that the correlation energy of 
intermediate-frequency fluctuations, which appear only in the crossover regime,
is of the order of 0.5--1~$\mu$eV, also smaller than $k_{B}T$.  For
low-frequency fluctuations, the correlation energy varies from about 
25~$\mu$eV in the strongly insulating regime, to 45~$\mu$eV in the transition 
region (assuming $D(E)\sim10^{14}$cm$^{-2}$eV$^{-1}$), down to 5~$\mu$eV at the
highest $\sigma$.  Therefore, it appears that it has a maximum in the 
crossover regime.  It is comparable to or larger than $k_{B}T$.  In
addition to having obtained qualitative agreement with the expected
behavior of a Coulomb interaction energy, we can also make a reliable
quantitative comparison in the most insulating regime, where $U$ in our devices
is reduced only by screening by the metallic gate~\cite{scr}.  Then
$U\sim 25~\mu$eV corresponds to a reasonable distance of 0.3~$\mu$m,
comparable to the mean separation ($\sim 0.8~\mu$m) of localized states lying 
within
an energy of a typical level spacing of $\mu$.  Using the same expression for 
$U$ in the case of intermediate-frequency fluctuations, $U\sim 1~\mu$eV 
gives a similar distance of 1~$\mu$m.  Even though, strictly speaking, this
estimate is not quite accurate in the transition region, it could indicate 
that these fluctuations reflect some
more subtle properties of the Coulomb interactions very close to the
MIT,  -- perhaps related to the beginning of overlap between chains of states
or more metallic islands --
as a result of the competition of several important length scales.

In addition to three distinct correlation voltages, we have
identified two characteristic voltage scales, $v_{0}$ and $v_{c}$.  $v_{c}$ is
a measure of the width of the intermediate-frequency regime, observable most
clearly in the crossover region.  In 
energy units, $v_{c}$ varies from about 15~$\mu$eV at low $\sigma$, to 
0.6~$\mu$eV in the transition region, up to about 1.5~$\mu$eV at high 
$\sigma$.  This indicates the existence of
a minimum of this characteristic energy as $\sigma$ is varied.  Our current 
data also show a tendency for the minimal value of $v_{c}$ in Fig.~\ref{third}
to go to zero as $T$ is lowered but more detailed measurements of the 
temperature dependence would be useful.
On general grounds, one expects that the characteristic length scales, such as
the localization length, will diverge at the MIT, and that the associated
characteristic energy will go to zero.  Our results for $v_{c}$ thus strongly
support the existence of a MIT in this 2D system.

In summary, we have analyzed the power spectra of fluctuations $\delta\ln G$
as a function of $V_{g}$.  
None of the non-interacting theories, which might be applicable to our
experiment~\cite{comment} within a limited range of $\sigma$, provide an
adequate description of our findings,
either because of the failure to predict the correct
shape of the low-frequency part of the spectrum~\cite{rr,wl} or because of the
failure to describe the observed temperature dependence of the correlation 
voltages~\cite{glazman,larkin}.  But most importantly, general considerations
based on non-interacting models~\cite{gang} suggest monotonic behavior of all
characteristic energy scales as a function of $n_{s}$.  In this paper, we
present a striking finding of a strong {\em nonmonotonic} behavior of several
energy scales, which may be understood by considering the effects of Coulomb
interactions.  Our experiment makes it absolutely clear that a theory
that treats both disorder and electron-electron interactions on an equal
footing is required in order to describe the MIT in this 2D system.

The authors are grateful to B. L. Altshuler, V. Dobrosavljevi\'{c}, A. B. 
Fowler, and B. B. Mandelbrot for useful discussions.  This work was supported 
by NSF Grant No. DMR-9510355.


\begin{references}

\bibitem{Dobro} V. Dobrosavljevi\'{c} and G. Kotliar, to be published in
Phys. Rev. Lett. (1997); 
D. Belitz and T. R. Kirkpatrick, Rev. Mod. Phys.~{\bf 66}, 261 (1994).

\bibitem{gang} E. Abrahams, P. W. Anderson, D. C. Licciardello, and
T. V. Ramakrishnan, Phys. Rev. Lett.~{\bf 42}, 673 (1979).

\bibitem{weakloc} D. J. Bishop, D. C. Tsui, and R. C. Dynes, Phys. Rev.
Lett.~{\bf 44}, 1153 (1980); M. J. Uren, R. A. Davies, M. Kaveh, and
M. Pepper, J. Phys. C~{\bf 14}, 5737 (1981).

\bibitem{Krav} S. V. Kravchenko, G. V. Kravchenko, J. E. Furneaux, V. M.
Pudalov, and M. D'Iorio, Phys. Rev. B~{\bf 50} 8039 (1994);
S. V. Kravchenko, Whitney E. Mason, G. E. Bowker, J. E. 
Furneaux, V. M. Pudalov, and M. D'Iorio, Phys. Rev. B~{\bf 51}, 7038 (1995);
S. V. Kravchenko, D. Simonian, M. P. Sarachik, Whitney Mason, and J. E.
Furneaux, Phys. Rev. Lett.~{\bf 77}, 4938 (1996).

\bibitem{AFS} T. Ando, A. B. Fowler, and F. Stern, Rev. Mod. Phys.~{\bf 54},
437 (1982).

\bibitem{Pop} Dragana Popovi\'{c}, A. B. Fowler, and S. Washburn, Phys. Rev.
Lett.~{\bf 67}, 2870 (1991).

\bibitem{glazman} L. I. Glazman and K. A. Matveev, Zh. Eksp. Teor. Fiz.
{\bf 94}, 332 (1988) [Sov. Phys. JETP~{\bf 67}, 1276 (1988)].

\bibitem{CB} H. van Houten, C. W. J. Beenakker, and A. A. M. Staring, in
{\em Single charge tunneling: Coulomb blockade phenomena in nanostructures},
ed. by Hermann Grabert and Michel H. Devoret (Plenum Press, New York, 1992).

\bibitem{Savchenko} V. V. Kuznetsov, A. K. Savchenko, M. E. Raikh, L. I. 
Glazman, D. R. Mace, E. H. Linfield, and D. A. Ritchie, Phys. Rev. B~{\bf 54},
1502 (1996).

\bibitem{Gleb} Gleb Finkelstein, Hadas Shtrikman, and Israel Bar-Joseph,
Phys. Rev. Lett.~{\bf 74}, 976 (1995).

\bibitem{Lee} J. G. Massey and Mark Lee, Phys. Rev. Lett.~{\bf 77}, 3399
(1996).

\bibitem{Efetov} K. B. Efetov, Phys. Rev. Lett.~{\bf 74}, 2299 (1995).

\bibitem{scr} L. D. Hallam, J. Weis, and P. A. Maksym, Phys. Rev. B~{\bf 53},
1452 (1996).

\bibitem{comment} We have analyzed fluctuations of $\ln G$ even though,
strictly speaking, an analysis of $\delta G$ would have been more appropriate
at higher values of $\sigma$.  This, however, affects only the results for 
variance.

\bibitem{rr} M. \'{E}. Raikh and I. M. Ruzin, Zh. Eksp. Teor. Fiz. {\bf 92},
2257 (1987) [Sov. Phys. JETP {\bf 65},1273 (1987)].

\bibitem{wl} P. A. Lee, A. Douglas Stone, and H. Fukuyama, Phys. Rev. B~{\bf 
35}, 1039 (1987).

\bibitem{larkin} A. I. Larkin and K. A. Matveev, Zh. Eksp. Teor. Fiz. {\bf 93},
1030 (1987) [Sov. Phys. JETP {\bf 66}, 580 (1987)].

\end{references}
\end{document}